\advance\topmargin by 2.0in
\documentclass[aps,prl,twocolumn,floatfix]{revtex4}
\usepackage{graphicx}
\usepackage{epsfig}
\usepackage{amsmath}

\begin{document}


\title{Phenomenology of One-Dimensional Quantum Liquids Beyond the Low-Energy
Limit}

\author{Adilet Imambekov and Leonid I. Glazman }
\affiliation{Department of Physics, Yale University, New Haven,
Connecticut, USA, 06520}

\date{\today}

\begin{abstract}
We consider zero temperature behavior of dynamic response
functions  of 1D systems near edges of support in momentum-energy
plane $(k, \omega).$ The description of the singularities of
dynamic response functions  near an edge $\varepsilon(k)$ is
given by the effective Hamiltonian of a mobile impurity moving in
a Luttinger liquid.  For Galilean-invariant systems, we relate the
parameters of such an effective Hamiltonian to the properties of
the function $\varepsilon (k).$ This allows us to express the
exponents which characterize singular response functions of
spinless bosonic or fermionic liquids in terms of
$\varepsilon(k)$ and Luttinger liquid parameters for any $k.$
For an antiferromagnetic Heisenberg spin$-1/2$ chain in a zero
magnetic field, $SU(2)$ invariance fixes the exponents from purely
phenomenological considerations.

\end{abstract}

\maketitle

\def\be{\begin{equation}}
\def\ee{\end{equation}}
\def\bea{\begin{eqnarray}}
\def\eea{\end{eqnarray}}

One of the central problems in condensed matter theory is the
development of an effective phenomenological description of
complicated many-body systems, the microscopic details of which
are often not known. The description of {\it low energy}
properties of interacting electrons in normal metals, for
example, is provided by the theory of Fermi
liquid~\cite{Nozieres}, while in one-dimensional systems
Luttinger liquid (LL) theory~\cite{Haldane,Giamarchibook,GNT}
plays a similar role. These phenomenological theories do not rely
on specific microscopic details, but predict certain low energy
properties of many-body systems in terms of few measurable
parameters. For example, LL theory assumes a linear spectrum of
low energy excitations, and  relates long-range behavior of
correlation functions  to the dimensionless LL parameter $K.$
However, the linear spectrum  approximation is not sufficient for
finding dynamic response functions (DRFs) even in low energy
limit~\cite{universal}. In this Letter we show, that for a wide
class of 1D systems one can phenomenologically predict certain
properties of DRFs in terms of  other measurable quantities even
beyond the low energy limit.

  The nonlinearity of the excitation spectrum affects transport phenomena, such as Coulomb drag~\cite{drag} and
momentum-resolved tunneling of electrons~\cite{Yacoby_science}
between quantum wires. In addition, neutron scattering on spin
chains~\cite{NS}, ARPES on quasi-1D materials~\cite{ARPES},   and
photoemission spectroscopy~\cite{photo_Jin} of 1D ultracold
atomic gases directly measure DRFs, and are not limited to low
energies. Evaluation of DRFs of 1D quantum systems with generic
excitation spectrum is also a test bed for rapidly developing
methods of numerical simulations of many-body
dynamics~\cite{DMRG,Affleck2007}. Recently some progress was
achieved in the analytical treatment of correlation functions
beyond linear spectrum
approximation~\cite{universal,Affleck2007,Pustilnik2006Fermions,Khodas2006Fermions,Pereira_bosons,ImambekovGlazman_PRL,CheianovPustilnik,ZCG,Arikawa,Carmelo,PustilnikCS,KG,bratskayamogila}.
The majority of analytical work however relied on solutions of
microscopic models, using perturbation theory
methods~\cite{Pustilnik2006Fermions,Khodas2006Fermions} or
integrability of models with specially tuned
parameters~\cite{Affleck2007,Pereira_bosons,ImambekovGlazman_PRL,CheianovPustilnik,ZCG,Arikawa,Carmelo,PustilnikCS}.
In contrast, the phenomenology developed in this Letter  does not
require any special property of the underlying microscopic
interaction, while it provides relations between different
experimentally observable quantities, such as the energy spectrum
and the exponents of DRF singularities; see  Eqs.~(\ref{Akwpower})
and~(\ref{deltaseq}).

\begin{figure}
\includegraphics[width=8cm]{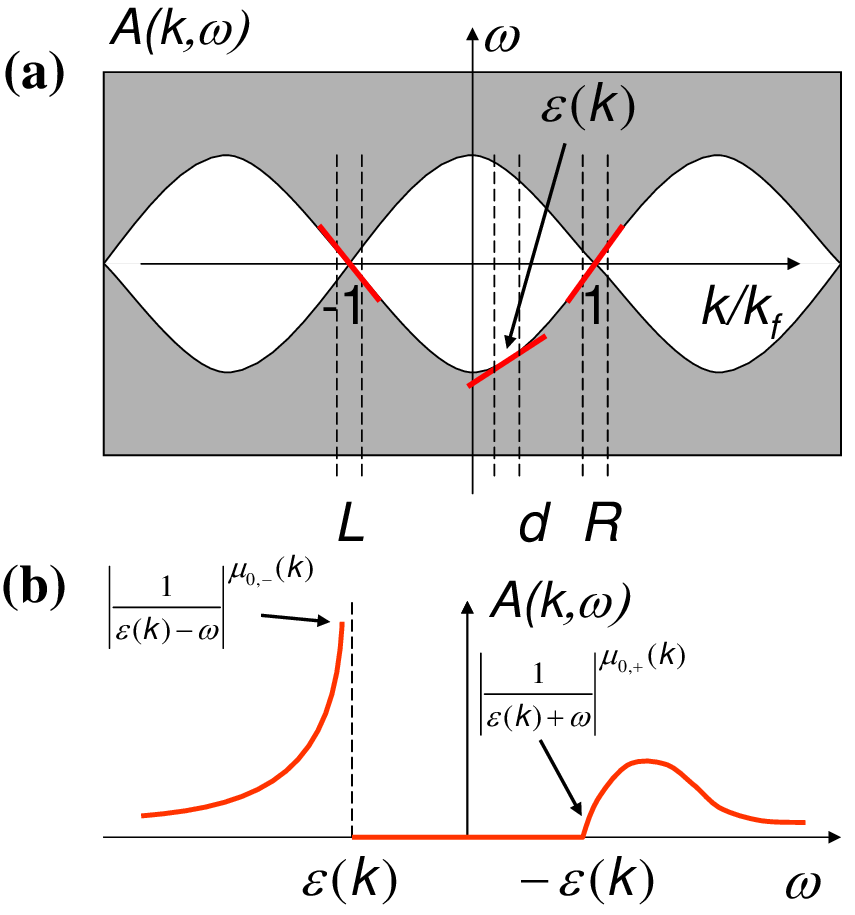}
\caption{\label{Fig1} (Color online) (a) Spectral function $A(k,
\omega)$ for spinless fermions in momentum-energy plane at $T=0.$
Shaded areas indicate the regions where $A(k, \omega)\neq 0,$ and
$\varepsilon(k)$ is the edge of the support in the basic region
$|k|<k_f,\omega<0,$ where $k_f$ is Fermi momentum. Edges in other
regions can be obtained from  $\varepsilon(k)$ by combinations of
shifts and inversions. (b) A sketch of constant $k$ scan of the
spectral function for $|k|<k_f.$ Singularity of $A(k, \omega)$
near $\omega\approx \varepsilon(k)$ can be described using
three-subband model of a mobile impurity moving in a Luttinger
liquid, Eqs.~(\ref{H0})-(\ref{Akwrepr}), and the answer is given
by Eqs.~(\ref{Akwpower}),(\ref{deltaseq}). }
\end{figure}

For spinless fermionic Galilean-invariant systems, the DRFs have a
sharp edge of support $\varepsilon(k)$ in the thermodynamic limit
at $T=0;$ see Fig. \ref{Fig1}.  Hamiltonian describing
singularities of DRFs, such as dynamic structure factor (DSF)
$S(p,\omega)$ and spectral function $A(p,\omega)$ [defined below
by Eqs.~(\ref{DSF})-(\ref{Gdef})], is the effective Hamiltonian of
a mobile impurity moving in a
LL~\cite{Affleck2007,Pustilnik2006Fermions,Khodas2006Fermions,impurity,Balents2000};
see Eqs.~(\ref{H0})-(\ref{Hint}) below.
Singularities of DRFs at the edges of support are the main subject
of this Letter. We show that  their exponents for
Galilean-invariant systems with interactions decaying faster than
$\propto 1/x$ can be expressed as functions of $\varepsilon(k)$
and LL parameters. Phenomenological considerations allow us also
to resolve the discrepancy~\cite{Affleck2007,CheianovPustilnik}
regarding antiferromagnetic spin$-1/2$ XXZ model in zero magnetic
field in favor of Ref.~\cite{Affleck2007}.

We are  interested in the zero temperature DSF \bea S (k,\omega) =
\int\!dx\,dt\,e^{i (\omega t-kx)}\, \bigl \langle \rho (x, t)
\rho (0, 0) \bigr \rangle, \label{DSF} \eea
 and spectral function
$ A(k,\omega)= -\frac1{\pi}{\rm Im }G(k, \omega)\, {\rm
sign}\omega, $ where Green's function $G(k,\omega)$ is defined by
\cite{AGD} \bea G(k, \omega)= -i \int\int dx dt e^{i (\omega t
-kx )} \bigl \langle T\left(\Psi(x,t)\Psi^{\dagger}(0,0)\right)
\bigr \rangle . \label{Gdef}\eea Here $\Psi(x,t)$ and $\rho (x,
t)$ are annihilation and density operators, respectively, and $T$
denotes time ordering. Energy $\omega$ is measured from chemical
potential, so $A(k,\omega)$ for $\omega>0 \;(\omega<0)$ describes
the response of the system to an addition of an extra particle
(hole).

To be specific, we  first discuss singularities of fermionic
spectral function in the region $|k|<k_f,\omega<0,$ where $k_f$
is Fermi momentum. Singularity can be described
\cite{Pustilnik2006Fermions,Khodas2006Fermions,Affleck2007} by
the effective Hamiltonian \bea H_0= \frac{v}{2\pi}\int dx\;\left[
K (\nabla \theta)^2 +
\frac1{K}(\nabla \phi)^2\right], \label{H0} \\
 H_d=\int dx\; d^{\dagger} (x) \left[\varepsilon(k)-i v_d \frac{\partial}{\partial
 x}\right]d(x),\label{Hd}\\
 H_{int}=\int dx \left[ V_{R} \rho_R(x)+V_{L}\rho_L(x)\right] \rho_d(x)\nonumber\\
 =\int dx \left( V_{R} \nabla \frac{\theta-\phi}{2\pi}-V_{L}\nabla\frac{ \theta+\phi}{2\pi}\right) d (x)
 d^{\dagger}(x).\label{Hint}
 \eea
Here $v$ is the  sound velocity, and fields $\theta$ and $\phi$
describe low energy  excitations
and have a commutation relation $[\phi(x),\nabla \theta(x')]=i
\pi \delta(x -x')$ (we use the notation of Ref.
\cite{Giamarchibook}). Operator $d(x)$ creates a mobile hole of
momentum $k$ and velocity $ v_d= \partial \varepsilon(k)/\partial
k,$ and operator $
 \rho_d(x)=d (x) d^{\dagger}(x)$ is the  hole density.
 In terms of hole operator $d(x,t),$
The singular part of  the spectral function
near $\varepsilon(k)$  is given by\bea A(k,\omega)\propto \int dx
dt e^{i\omega t} \langle d^{\dagger}(x,t) d(0,0)
\rangle_{H_0+H_d+H_{int}}.\label{Akwrepr} \eea Canonical
transformation $ \phi=\tilde\phi \sqrt{K} ,\theta=\tilde\theta
/\sqrt{K}$ diagonalizes $H_0,$ while  the term $H_{int}$ can be
removed~\cite{Balents2000} by unitary transformation
$U^{\dagger}(H_0+H_d+H_{int})U,$ where \be U^{\dagger}=e^{-i\int
dx \left\{\frac{\delta_+(k)}{2\pi}[\tilde \theta(x)-\tilde
\phi(x)] -\frac{\delta_-(k)}{2\pi}[\tilde \phi(x)+\tilde
\theta(x)]\right\}d(x)d^{\dagger}(x) }. \nonumber \ee Momentum
dependent phase shifts $\delta_+(k),\delta_-(k)$  are related to
the parameters of $H_{int}$ as \bea
\left(V_{L}-V_{R}\right)/\sqrt{K}=- \delta_{-}(k)(
v_d+v)+\delta_{+}(k)( v_d-v),\label{Vviadelta1}\\
\left(V_{L}+V_{R}\right)\sqrt{K}=- \delta_{-}(k)(
v_d+v)-\delta_{+}(k)(v_d-v).\label{Vviadelta2} \eea
Calculating $U^{\dagger}d(x)U$ together with Eq.~(\ref{Akwrepr}),
one obtains \bea A(k,\omega) \propto
\theta(\varepsilon(k)-\omega)\left|\frac1{\varepsilon(k)-\omega}\right|^{1-\left[\frac{\delta_+(k)}{2\pi}\right]^2-\left[\frac{\delta_-(k)}{2\pi}\right]^2}.\label{Akwpower}
\eea
To obtain phase shifts, one needs to fix $V_R$ and $V_L$ in
Eq.~(\ref{Hint}). We relate  $V_R$ and $V_L$ to $\varepsilon(k)$
by calculating  in two ways the shift of the position of the edge
under uniform density and current variations.

Uniform density variation $\delta \rho$ results in  a finite
expectation value $\langle \nabla \varphi \rangle=-\pi
\delta\rho.$ Evaluating
 the shift of $\varepsilon(k)$ in two ways, we obtain  \bea -\frac{V_R + V_L}{2}= \frac{\partial
\varepsilon(k)}{\partial\rho} + \frac{\partial \mu}{\partial
\rho}=\frac{\partial \varepsilon(k)}{\partial\rho} +
\frac{v\pi}{K}. \label{univV2} \eea The left-hand side follows
from the effective Hamiltonian given by
Eqs.~(\ref{H0})-(\ref{Hint}), while the right-hand side follows
from the evaluation of the edge position from its thermodynamic
definition [taking into account that energy $\varepsilon(k)$ is
measured with respect to chemical potential $\mu$].

Uniform current through the system results in a finite value of
$\langle \nabla \theta \rangle,$ which for Galilean-invariant
systems corresponds to a motion with constant velocity $
u=\langle \nabla \theta \rangle/m,$ where $m$ is the bare mass of
the constituent particles. Then following  the argument of
Refs.~\cite{Baym,KG}, one can use Galilean invariance to evaluate
the change of $\varepsilon(k).$ Comparing it with the change
evaluated using Eqs.~(\ref{H0})-(\ref{Hint}) leads to \bea
\frac{V_L - V_R}{2\pi}=\frac{k}{m}- \frac{\partial
\varepsilon(k)}{\partial k}. \label{univV1}\eea

Combining now Eqs. (\ref{Vviadelta1}) and (\ref{Vviadelta2}) with
Eqs. (\ref{univV2}) and (\ref{univV1}), we obtain the central
result of this Letter
 \bea
\frac{\delta_{\pm}(k)}{2\pi}=
\frac{\frac1{\sqrt{K}}\left(\frac{k}{m}-\frac{\partial
\varepsilon(k)}{\partial k}\right) \pm \sqrt{K}\left(\frac{1}{\pi}
\frac{\partial \varepsilon(k)}{\partial \rho} +
\frac{v}{K}\right)}{2\left(\pm \frac{\partial
\varepsilon(k)}{\partial k} -v\right)}. \label{deltaseq}\eea  On
the basis of Galilean invariance, it establishes a
model-independent phenomenological relation between the edge
position $\varepsilon(k)$ and other measurable quantities, such as
the exponent of spectral function, Eq.~(\ref{Akwpower}).
Even for usual LL theory Galilean-invariant systems are special.
For them, LL parameter $K$
can be expressed~\cite{Haldane} as a renormalization of sound
velocity  compared to Fermi velocity $v_f$  in the absence of
interactions, $K=v_f/v=\pi \rho/(m v).$ Our results are  a
generalization of the special properties of Galilean systems
beyond low energy limit.

While Eq.~(\ref{univV1}) doesn't hold on a lattice,
Eq.~(\ref{univV2}) still works, and will be used below for the XXZ
model. One can formulate an analog of Eq.~(\ref{univV1}) for LLs
on lattices using the derivative of $\varepsilon(k)$ with respect
to total flux through the system under periodic boundary
conditions. Energies are easier to evaluate numerically than
correlation functions, so our results can be used as a benchmark
for numerical methods for evaluation of DRFs.

Away from the basic region $|k|<k_f,\omega<0,$ positions of edges
can be obtained from $\varepsilon(k)$ in the basic region by a
combination of inversions and shifts. States which define the
positions of the edges are given by a hole and excitations near
Fermi points. Exponents of divergences can also be obtained using
the three-subband model given by Eqs. (\ref{H0})-(\ref{Hint}), and
here we only summarize the results.

For the spectral function momentum $k$ in the region
$(2n-1)k_f<k<(2n+1)k_f,$ hole momentum equals $k_n = k - 2 n k_f.$
Near the edges for $\omega>0 (\omega<0),$ the spectral function
is defined by \bea A(k,\omega) \propto \theta( \varepsilon(k_n)\pm
\omega)\left|\frac1{\omega\pm\varepsilon(k_n)}\right|^{\mu_{n,\pm}(k)},\\
\mu_{n,\pm}(k)=
1-\frac12\left(2n\sqrt{K}-\frac{\delta_+(k_n)+\delta_-(k_n)}{2\pi}\right)^2\nonumber
\\-\frac12\left(\frac{1\pm 1}{\sqrt{K}}+\frac{\delta_+(k_n)-\delta_-(k_n)}{2\pi}\right)^2. \label{munpm}\eea

DSF $S(k,\omega)$ is non-vanishing only for $\omega>0,$ and for $
2n k_f<k< 2(n+1) k_f $ hole momentum is given by $
k_n^*=(2n+1)k_f-k.$ The exponent of DSF $\mu_{n}(k)$ is defined by
\bea S (k,\omega) \propto  \theta(\omega+
\varepsilon(k^*_n))\left|\frac1{\omega+\varepsilon(k^*_n)}\right|^{\mu_{n}(k)},\\
\mu_{n}(k)=
1-\frac12\left[(2n+1)\sqrt{K}+\frac{\delta_+(k^*_n)+\delta_-(k^*_n)}{2\pi}\right]^2\nonumber\\
-\frac12\left(\frac1{\sqrt{K}}+\frac{\delta_+(k^*_n)-\delta_-(k^*_n)}{2\pi}\right)^2.
\label{mun} \eea

For bosons, spectral function has divergences at
$\mp\varepsilon(k^*_n)$  for $\omega>0 (\omega<0),$ and exponents
equal \bea \mu^b_{n,\pm}(k)=
1-\frac12\left[(2n+1)\sqrt{K}-\frac{\delta_+(k^*_n)+\delta_-(k^*_n)}{2\pi}\right]^2\nonumber
\\-\frac12\left(\frac{1\pm 1}{\sqrt{K}}+\frac{\delta_+(k^*_n)-\delta_-(k^*_n)}{2\pi}\right)^2. \;\;\label{mubnpm}\eea

Let us now discuss several cases where one can explicitly check
our phenomenological predictions. The shift of the position of
$\varepsilon(k)$ can be evaluated using perturbation theory in
interaction strength for any momenta, and predictions following
from our theory coincide with results of
Refs.~~\cite{Pustilnik2006Fermions,Khodas2006Fermions}. By using
approximation $\varepsilon(k)\approx v(k-k_f)$ for any
interaction strength in the vicinity of the right Fermi point,
from Eqs. (\ref{Vviadelta2}) and (\ref{univV2}) one can recover
the universal phase shift ~\cite{universal}
 \bea
\frac{\delta_{-}(k_f-0)}{2\pi}=\frac{1}{2\sqrt{K}}-\frac{\sqrt{K}}{2}
\label{delta-},
 \eea
which holds irrespective of Galilean invariance. One can also
obtain $\delta_{+}(k_f-0)$ from Eq.~(\ref{deltaseq}).
 For that, one has to use
the expansion $\varepsilon(k)\approx v(k-k_f)+(k-k_f)^2/(2m_*),$
and the expression for $1/m_*$ obtained in
Ref.~\cite{Pereira_bosons}, which is valid for Hamiltonians with
interactions decaying faster than $\propto 1/x^2.$ For
Galilean-invariant systems, it simplifies to $
1/m_*=[\sqrt{K}/(2\pi)]\partial{v}/\partial{n} +1/(2 m \sqrt{K}),$
and after some simple algebra with Eq.~(\ref{deltaseq}) one
reproduces universal phase shift~\cite{universal}
$\delta_{+}(k_f-0)/(2\pi)=1-1/(2\sqrt{K})-\sqrt{K}/2.$ One can
also explicitly check, that  exponents
 for Lieb-Liniger~\cite{Lieb} and
Calogero-Sutherland~\cite{BM} models evaluated from their
excitation spectra reproduce the results of
Refs.~\cite{ImambekovGlazman_PRL} and
~\cite{Khodas2006Fermions,PustilnikCS}, respectively.

The crucial step in the calculation of the exponents is the
identification of the spectral function $A(k,\omega)$ defined in
terms of constituent particles, Eq.~(\ref{Gdef}), with the
correlation function of operator $d$ in Eq.~(\ref{Akwrepr}).
Comparison with solvable cases above shows that such
identification indeed holds in the vicinity of Fermi points for
any interactions, as well as for any momentum for weak
interactions. While we cannot prove that it holds for any
strongly interacting Galilean-invariant system, we expect it to
be valid as long as the position of the edge satisfies \bea
\left|\frac{\partial\varepsilon(k)}{\partial k}\right| <v \;\;
\mbox{for} \; |k|<k_f, \label{strong_ineq} \eea and interactions
decay faster than $\propto 1/x.$ Equation~(\ref{strong_ineq})
guarantees that phases in Eq.~(\ref{deltaseq}) are continuous
functions of momentum, and the state which corresponds to the edge
of the basic region of the spectral function support does not
contain particle-hole excitations near left or right Fermi points.

Let us now discuss how considerations of this Letter can resolve
a discrepancy between results of Refs.~\cite{Affleck2007} and
\cite{CheianovPustilnik} for correlations of the XXZ model in zero
magnetic field. In our notations, these references predict
$\delta^{PWA}_{-}(k)/(2\pi)=-\delta^{PWA}_{+}(k)/(2\pi)=(1/\sqrt{K}-\sqrt{K})/2$
and
$\delta^{CP}_{-}(k)/(2\pi)=-\delta^{CP}_{+}(k)/(2\pi)=(1-K)/2,$
respectively. Identification of these phase shifts was based on
the analysis of finite size corrections to energies obtained from
the exact solution. Their interpretation  for the XXZ model in
zero magnetic field is ambiguous, since the half-filled lattice is
a special point for the Bethe Ansatz solution~\cite{KBI}. On the
other hand, our approach constrains phase shift via
$\varepsilon(k),$ which is well defined in the thermodynamic
limit, and resolves the discrepancy.

First, we note that $\delta^{PWA}_{-}(k)$ satisfies universal
relation given by Eq.~(\ref{delta-}), while $\delta^{CP}_{-}(k)$
does not. Second, Eqs. (\ref{Vviadelta1}),(\ref{Vviadelta2}) and
(\ref{univV2}) hold on the lattice for any $k,$ and one can
easily evaluate excitation spectrum of the XXZ model numerically
from the exact solution~\cite{KBI}. This way, we have verified
that $\delta^{PWA}_{\pm}(k)$ satisfy them, while
$\delta^{CP}_{\pm}(k)$ do not.  Third, one can use $SU(2)$
invariance to independently derive results of
Ref.~\cite{Affleck2007} at the XXX point. The argument is very
similar to the reasoning which fixes LL parameter $K=1/2$ at
$XXX$ point~\cite{Giamarchibook, GNT} by requiring that long
distance asymptotes of $\langle S^{z}(x)S^{z}(0)\rangle $ and
$\langle S^{-}(x)S^{+}(0)\rangle$ coincide. But $SU(2)$ symmetry
also establishes a relation between spin DRFs $S^{zz}(k,\omega)$
and $S^{-+}(k,\omega)$ in entire momentum-energy plane, including
the edges of supports. There $S^{zz}(k,\omega)$ behaves as the sum
of different power laws (up to logarithmic corrections) \bea
S^{zz} (k, \omega) \propto
  \sum_l \left|\frac{1}{\omega -
 \varepsilon(k)}\right|^{\mu^{zz}_l}.\eea
 Different power laws appear because of the umklapp processes that
 are allowed on a half-filled lattice. $SU(2)$ symmetry implies that
 the same set of exponents should apply for $S^{-+}(k,\omega)$ as
 well.  These exponents can be evaluated in terms of
 $\delta_{\pm}(k)$ and $K$ for any $l,$ and the coincidence of two sets of exponents unambiguously fixes \bea \frac{\delta_{\pm}(k)}{2\pi}= \mp \frac1{2\sqrt{2}},\eea
 as in Ref.~\cite{Affleck2007} for $K=1/2.$ Full sequence of exponents is \bea
 \mu^{zz}_l = 3/4-(4l+1)^2/4. \label{musec}
 \eea
Note, that we did not use integrability in the argument for the
XXX model, so Eq.~(\ref{musec}) should apply for $SU(2)$ invariant
models with longer range interactions as well, if the spin chain
remains a gapless LL.

 To summarize, we have considered zero
temperature dynamic response functions of 1D systems near edges
of support in the momentum-energy plane. Continuous symmetries can
be used  to fix the exponents of power law divergences of dynamic
response functions near the edges. For spinless Galilean-invariant
systems of fermions or bosons, we have obtained phenomenological
expressions,
Eqs.~(\ref{deltaseq}),(\ref{munpm}),(\ref{mun}), and
(\ref{mubnpm}), which establish model-independent relations of
the exponents of dynamic response functions to the position of
the edge of support $\varepsilon(k).$  For a spin$-1/2$
anitferromagnetic Heisenberg chain in zero magnetic field,
$SU(2)$ symmetry dictates exponents given by Eq.~(\ref{musec})
for all momenta regardless of the interaction range.

 We thank  A. Kamenev and A. Lamacraft for useful discussions, and NSF Grant DMR-0754613 for support.


\end{document}